\documentclass[prl
,twocolumn
,floatfix
,showpacs
,preprintnumbers
,amsmath
,amssymb
,superscriptaddress]{revtex4}

\usepackage[T1]{fontenc}
\usepackage{graphicx}
\usepackage{dcolumn}
\usepackage{bm}
\usepackage{color}
\usepackage{ulem}

\begin{document}
\title{Crackling dynamics in material failure as the signature of a self-organized dynamic phase transition}
\author{D. Bonamy}
\email{daniel.bonamy@cea.fr}
\affiliation{CEA, IRAMIS, SPCSI, Grp. Complex Systems $\&$ Fracture, F-91191 Gif sur Yvette, France}
\author{S. Santucci}
\affiliation{Fysisk Institutt, Universitetet i Oslo, P.O. Boks 1048
Blindern, N-0316 Oslo 3, Norway}
\affiliation{Present address:
Physics of Geological Processes, Universitetet i Oslo, P.O. Boks 1048
Blindern, N-0316 Oslo 3, Norway}
\author{L. Ponson}
\affiliation{CEA, IRAMIS, SPCSI, Grp. Complex Systems $\&$ Fracture, F-91191 Gif sur Yvette, France}
\affiliation{Present address: Division of Engineering and Applied Science, California Institute of Technology, Pasadena, CA 91125, USA}

\begin{abstract}
We derive here a linear elastic stochastic description for slow crack growth in heterogeneous materials. This approach succeeds in reproducing quantitatively the intermittent crackling dynamics observed recently during the slow propagation of a crack along a weak heterogeneous plane of a transparent Plexiglas block [Måløy {\it et al.}, PRL {\bf 96} 045501]. In this description, the quasi-static failure of heterogeneous media appears as a self-organized critical phase transition. As such, it exhibits universal and to some extent predictable scaling laws, analogue to that of other systems like for example magnetization noise in ferromagnets.
\end{abstract}

\date{\today}
\maketitle

Driven by both technological needs and the challenges of unresolved questions in fundamental physics, the effect of materials heterogeneities onto their failure properties has been extensively studied in the recent past (see \cite{Alava06_ap} for a recent review). So far, many efforts have been focusing on the morphology of fracture surfaces (see \cite{Bouchaud97_jpcm} for a review). In particular, crack surface roughness was recently shown to exhibit anisotropic morphological scaling features \cite{Ponson06_prl,Bonamy06_prl} that could be understood for brittle materials \cite{Bonamy06_prl}.

Here, we will focus our study on the dynamics of cracks. In heterogeneous materials under slow external loading, this propagation displays a jerky dynamics with seemingly random sudden jumps spanning over a broad range of length-scales \cite{directcrackling}. Such a complex dynamics also called ``Crackling Noise" \cite{Sethna01_nature} has been also suggested from the acoustic emission accompanying the failure of various materials \cite{acoustic} and the seismic behaviour accompanying earthquakes \cite{earthquakes} characterized by power-law energy distribution with an exponent around $1.3--1.5$. These distributions can be qualitatively reproduced in phenomenological models like Fiber Bundle Models (FBM) \cite{FBM} (resp. Random Fuse Models (RFM) \cite{RFM})which schematizes materials as a set of brittle fibers loaded in parallel with random failure thresholds and a rule for load redistribution after each failure event (resp. as a network of electrical fuse with constant resistance and randomly distributed threshold). However, these simple approaches yield to an exponent around  $1.9--2.5$ significantly higher than the experimental observations. Moreover, they rely on important simplifications which makes quantitative comparisons with experiments difficult.

We will demonstrate here that this crackling dynamics can be fully reproduced through a stochastic description rigorously derived from Linear Elastic Fracture Mechanics (LEFM) \cite{Lawn93_book} extended to disordered materials. In particular, this model succeeds in reproducing {\it quantitatively} the intermittent crackling dynamics observed recently during the steady slow crack growth along a weak heterogeneous plane within a transparent Plexiglas block \cite{Maloy06_prl}. In this description, quasi-static failure of heterogeneous brittle elastic media can be interpreted as a self-organized critical dynamic phase transition and - as such - exhibits universal behaviors. We will then show how one can use Universality and previous calculations performed on different systems belonging to the same universality class, here the motion of domain walls in disordered ferromagnets, to derive predictive laws for the failure of materials.

{\it Theoretical description:} LEFM is based on the fact that -- in an elastic medium under tensile loading -- the mechanical energy $G$ released as fracture occurs is entirely dissipated within a small zone at the crack tip. Defining the fracture energy $\Gamma$ as the energy needed to create two crack surfaces of unit area, under quasi-static condition, we assume that the local crack velocity $v$ is proportional to the excess energy locally released:
\begin{equation}
\frac{1}{\mu}v=G-\Gamma
\label{equ1}
\end{equation}
where $\mu$ is the effective mobility of the crack front.

In a homogeneous medium, $\Gamma=\Gamma_0$ is constant and an initially straight crack front will be translated without being deformed. LEFM allows the determination of the energy released $G$ in any geometry. In particular, for the experimental configuration chosen in \cite{Maloy06_prl}, considered here and depicted in Fig.1  where a crack grows at the interface between two elastic plates by lowering the bottom plate at constant velocity $V$, $G$ is given by \cite{Obreimoff30_rpsl}:
\begin{equation}
G^0(f(t))=\frac{1}{3}E\delta^3\frac{(\Delta_0+Vt)^2}{(c_0+f(t))^4}
\label{equ2}
\end{equation}
where $c_0+f(t)$ is the instantaneous crack length, $E$ the Young modulus, $\delta$ the thickness of the lowered plate, $\Delta_0$ the initial opening and $c_0$ the initial crack length. Then, considering a slow driving velocity such as $Vt\ll \Delta_0$, one can show from Eqs. \ref{equ1},\ref{equ2} that after a short transient regime, the crack front propagates at a constant velocity:
\begin{equation}
f(t)\simeq v_m t \; \mathrm{with} \; v_m=\frac{Vc_0}{2\Delta_0} \; \mathrm{and} \; c_0=\left(\frac{E\delta^3\Delta_0^2}{3\Gamma_0}\right)^{1/4}
\label{equ3}
\end{equation}

In a heterogeneous material, defects induce fluctuations in the local toughness: $\Gamma(x,y)=\Gamma^0(1+\eta(x,y))$ where the noise term $\eta(x,y)$ captures these fluctuations, and $x$ and $y$ denotes the coordinates in the propagation and crack front directions respectively. These fluctuations induce local distortions of the crack front $f(x,t)$, which in turn generate local perturbations in $G$ \cite{Schmittbuhl95_prl}. To linear order in $f$, one can show that \cite{Gao89_jam}:
\begin{equation}
\begin{array}{ll}
G(x,f(x,t))\approx & G^0\left(\left<f(x,t)\right>_x\right)\\
& + \frac{1}{2\pi}G^0\left(\left<f(x,t)\right>_x\right)\int_{-\infty}^{\infty}\frac{f(x',t)-f(x,t)}{(x'-x)^2}dx'
\end{array}
\label{equ4}
\end{equation}
And finally, by replacing this expression in Eq. \ref{equ1}, using the expression of $G^0$ for the homogeneous case, and by introducing the quantities defined in Eq. \ref{equ3}, we get:
\begin{equation}
\begin{array}{l}
\frac{1}{\mu\Gamma^0}\frac{\partial f}{\partial t} = F(t,\{ f\})+\frac{1}{2\pi}\int_\infty^\infty\frac{f(x',t)-f(x,t)}{(x'-x)^2}dx'-\eta(x,f(x,t))\\
\mathrm{where} \, F(t,\{ f\})=\frac{4}{c_0}\left( v_mt-<f(x,t)>_x\right)
\end{array}
\label{equ5}
\end{equation}
Strictly speaking, this equation describes the interfacial crack growth according to the geometry depicted in Fig. 1. But the very same equation -- with different prefactors in the expression of $F(t,\{ f\})=\frac{4}{c_0}\left( v_mt-<f(x,t)>_x\right)$ -- would have been obtained for the quasi-static stable crack growth of the in-plane component of the crack front within a three-dimensional solid independently of the tensile loading conditions and the system geometry.

\begin{figure}
\begin{center}
\includegraphics[width=0.9\columnwidth]{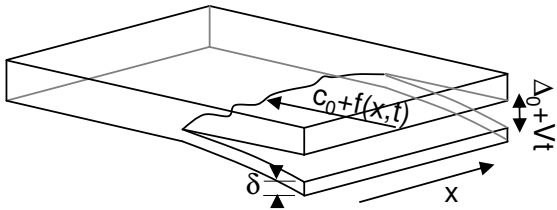}
\caption{Crack growth along the inhomogeneous interface between two elastic plates loaded according to the geometry used in \cite{Maloy06_prl}: Sketch and notations.}
\end{center}
\label{fig1}
\end{figure}

Variants of this equation with {\it constant} force $F$ have been extensively studied in the past to model crack propagation in solids \cite{Schmittbuhl95_prl}, but also to describe other systems as diverse as interfaces in disordered magnets \cite{Bertotti94_jap,Durin00_prl} or contact lines of liquid menisci on rough substrates \cite{Ertas94_pre,Rolley98_prl}. In this case, the interface remains stationary, pinned by the heterogeneities, unless this constant force $F$ is larger than a threshold value $F_c$. A key feature of these systems is that the so-called depinning transition at $F_c$ belongs to the realm of collective critical phenomena characterized by universal scaling laws \cite{NarayanNattermannKardar}. In particular, at $F_c$, the interface moves through scale free avalanches, both in space and time.

Equation \ref{equ5} denotes a rather different situation where the effective force $F(t,\{ f\})$ is not constant anymore, but given by the difference between the mean front position and the one that would have been observed within the homogeneous case: When $F(t,\{ f\}) <F_c$, the front remains pinned and $F(t,\{ f\})$ increases with time. As soon as $F(t,\{ f\}) > F_c$, the front propagates, $F(t,\{ f\})$ increases, and, as a consequence, $F(t,\{ f\})$ is reduced until the front is pinned again. This retroaction process keeps the crack growth close to the depinning transition {\it at each time} and, within the limit $v_m\rightarrow 0$ and $c_0\rightarrow \infty $  the system remains at the critical point during the whole propagation, as for self-organized-critical systems \cite{Bak87_prl}.

{\it Spatio-temporal intermittent dynamics --} Predictions of this stochastic description are now confronted to the experimental observations reported in \cite{Maloy06_prl}. Using a fourth order Runge-Kutta scheme, Eq. \ref{equ5} is solved for a front $f(x,t)$ propagating in a $1024\times 1024$ uncorrelated 2D random gaussian map $\eta(x,y)$ with zero average and unit variance [Note that the crackling dynamics statistics detailed thereafter do not depend on the precise distribution of $\eta$ as far as it remains short-range correlated]. The parameter $\mu\Gamma_0$ was set to unity while the two parameters $c_0$ and $v_m$ were varied from 2.5 to 250, and $10^{-2}$ to $5\times 10^{-1}$, respectively. In order to characterize the scaling features of the crack front local dynamics, we adopt the analysis procedure recently proposed in \cite{Maloy06_prl} and compute at each point $(x,y)$ of the recorded region the time $w(x,y)$ spent by the crack front within a small $1 \times 1$pixel$^2$ region  as it passes through this position. A typical gray-scale image of this so-called waiting time map are presented in Fig. 2a. The numerous and various regions of gray levels reflect the intermittent dynamics, and look very similar to those observed experimentally (Fig. 1b of \cite{Maloy06_prl}).  From the inverse value of waiting-time map, we compute a spatial map of the local normal speed velocity $v(x,y)=1/w(x,y)$ of the front as it passes through the location $(x,y)$. Avalanches are defined as clusters of velocities $v$ larger than a given threshold $v_c=C\bar{v}$ where $\bar{v}$ denotes the crack velocity averaged over both time and space within the steady regime. Their area is power-law distributed with an exponent $\tau_0=-1.65\pm0.05$ (Fig. 2c). These clusters exhibit morphological scaling features since their width, -- measured along the direction of crack growth --, is shown to go as a power-law with respect to their length -- measured along the direction of crack front -- with an exponent $H=0.65\pm0.05$ (Fig. 2d). All these results are in perfect agreement with those measured experimentally in \cite{Maloy06_prl}. Contrary to what conjectured in \cite{Maloy06_prl}, $H$ is {\it significantly different} from the value of the roughness exponent $\zeta \simeq 0.39$ expected for the interface at the depinning point of Eq. \ref{equ5} \cite{Schmittbuhl95_prl,roughnessexponent}. It is worth to mention that recent experimental results \cite{Santucciroughness08} report a roughness exponent $\zeta \simeq 0.35$ at large scale, in agreement with this theoretical prediction.
  
To complete the characterization of the avalanche statistics, we measure a new observable, the avalanche duration, -- defined as the difference between the times when the crack front leaves and arrives to the considered avalanche cluster (Fig. 2b). The distribution of the avalanche duration is plotted on Fig. 2e. For durations $D$ smaller than the average avalanche duration $<D>$, this distribution clearly depends on $v_m$, $c_0$ and the clipping threshold $C$ while for $D \gg <D>$ all the distributions seem to collapse onto a single power-law behaviour characterized by an exponent $\alpha_>=-2.1\pm0.2$ (Fig. 2e). Finally, the mean avalanche duration is found to go as a power law with the mean avalanche size, characterized by an exponent $\gamma=0.4\pm0.05$ (Fig. 2f).
\begin{figure}
\begin{center}
\includegraphics[width=0.8\columnwidth]{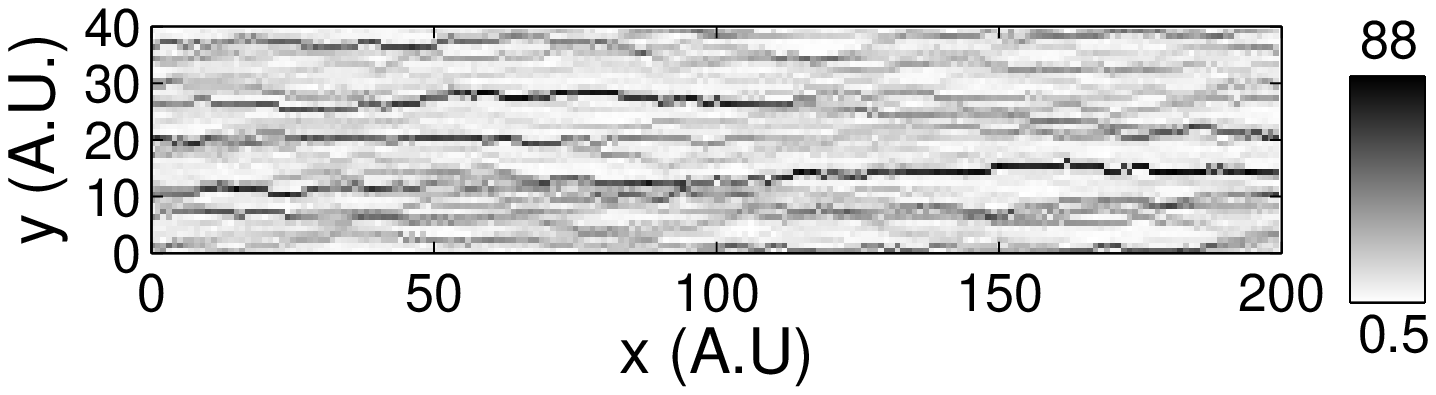}
\includegraphics[width=0.8\columnwidth]{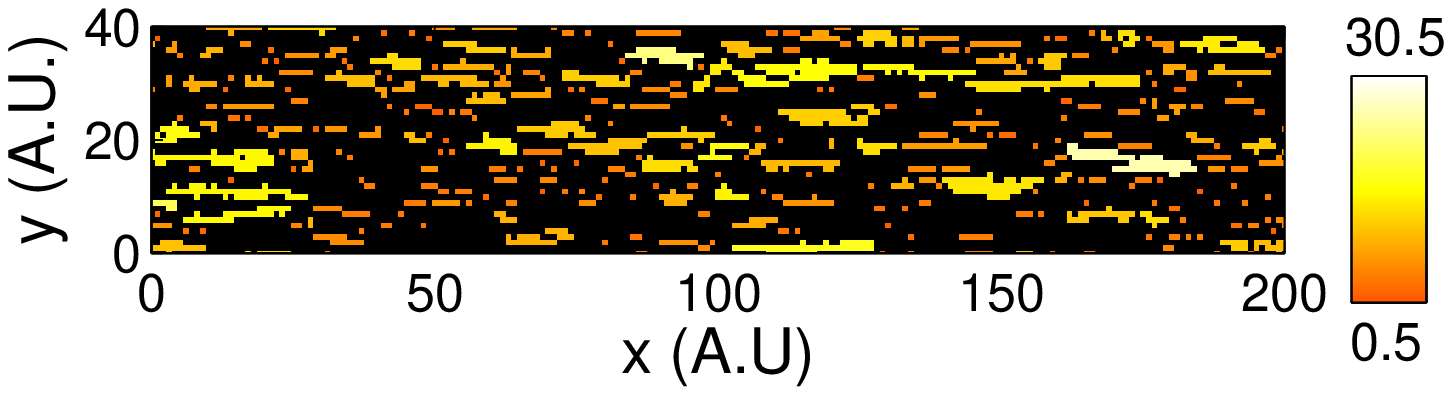}
\includegraphics[width=0.48\columnwidth]{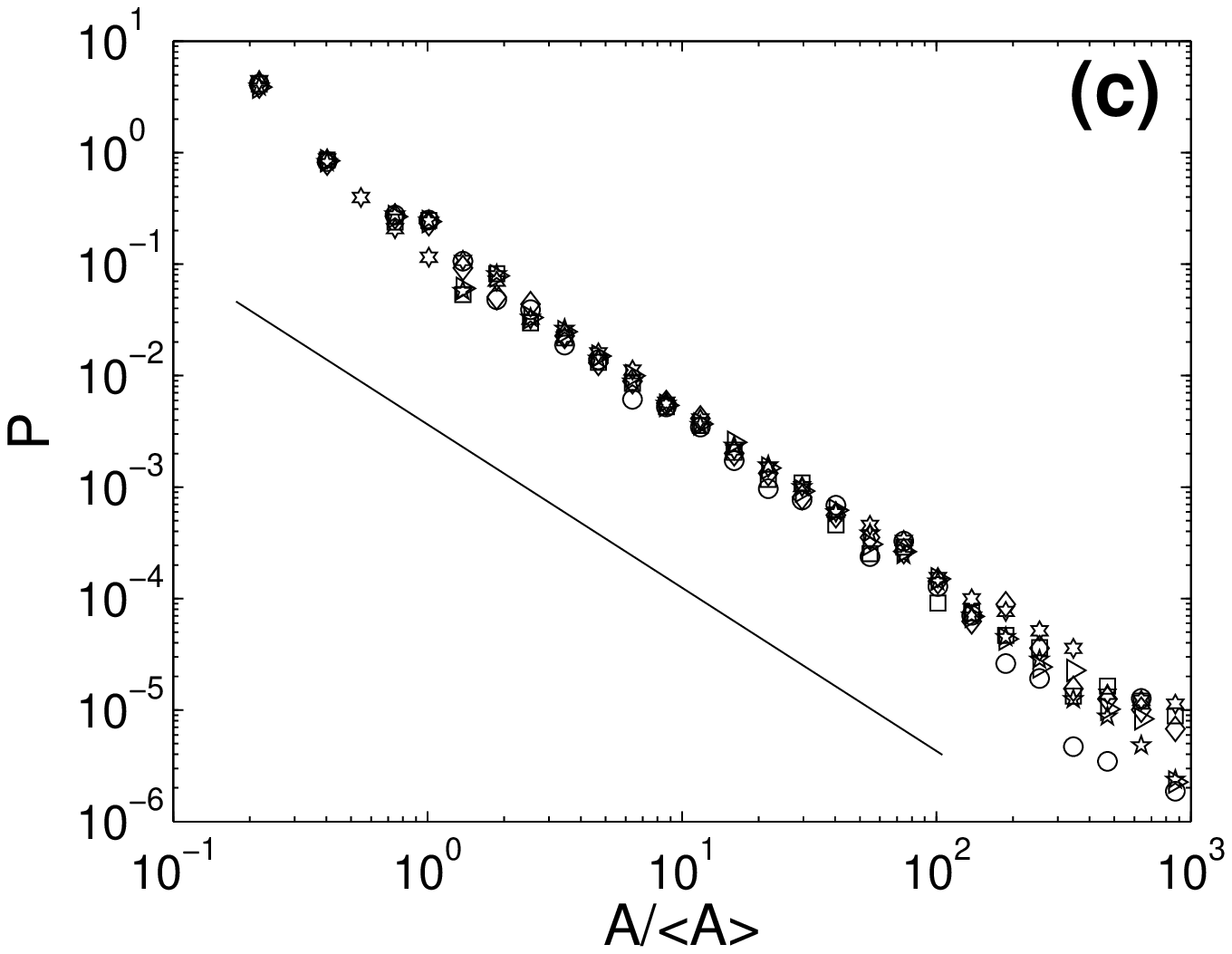}
\includegraphics[width=0.48\columnwidth]{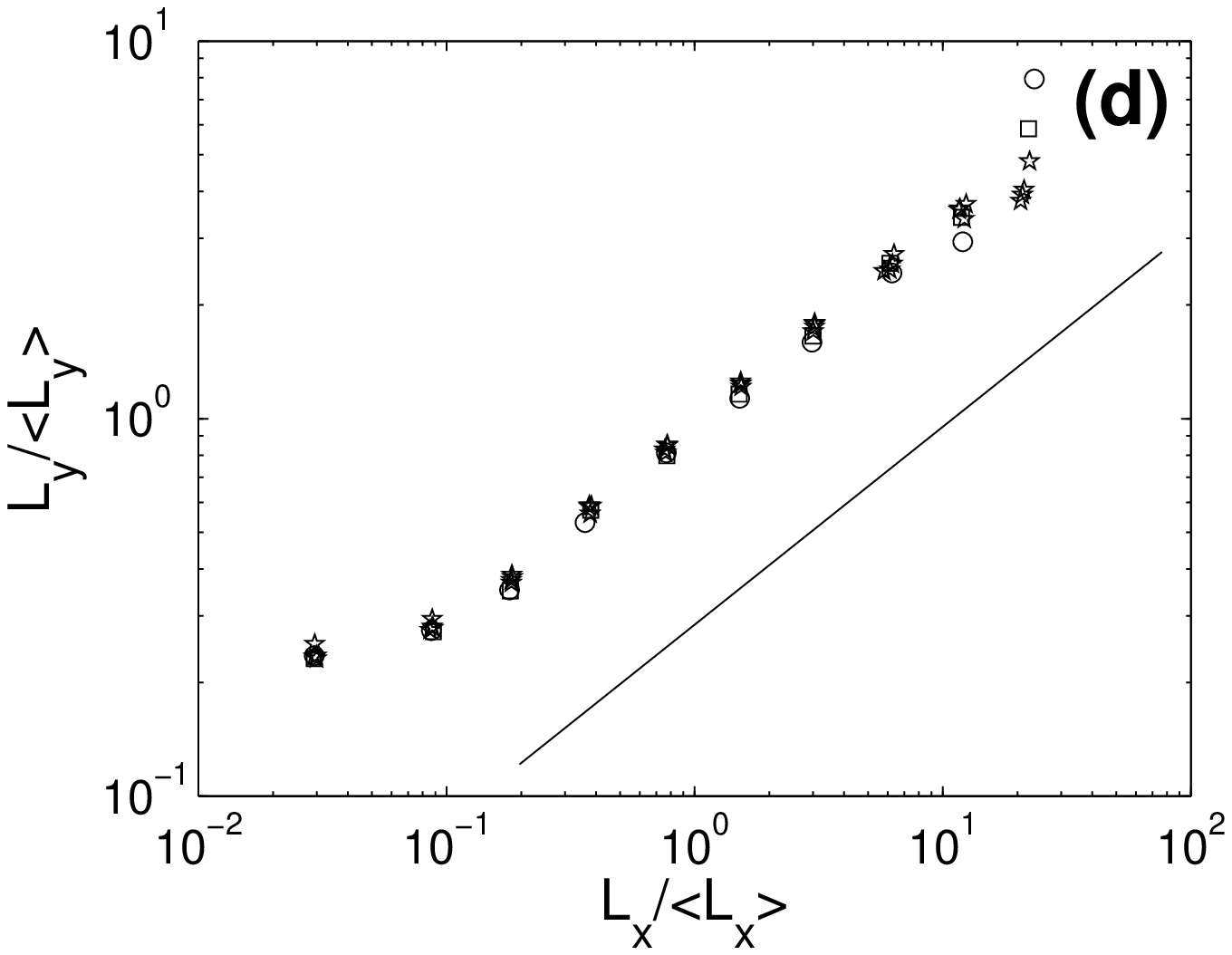}
\includegraphics[width=0.48\columnwidth]{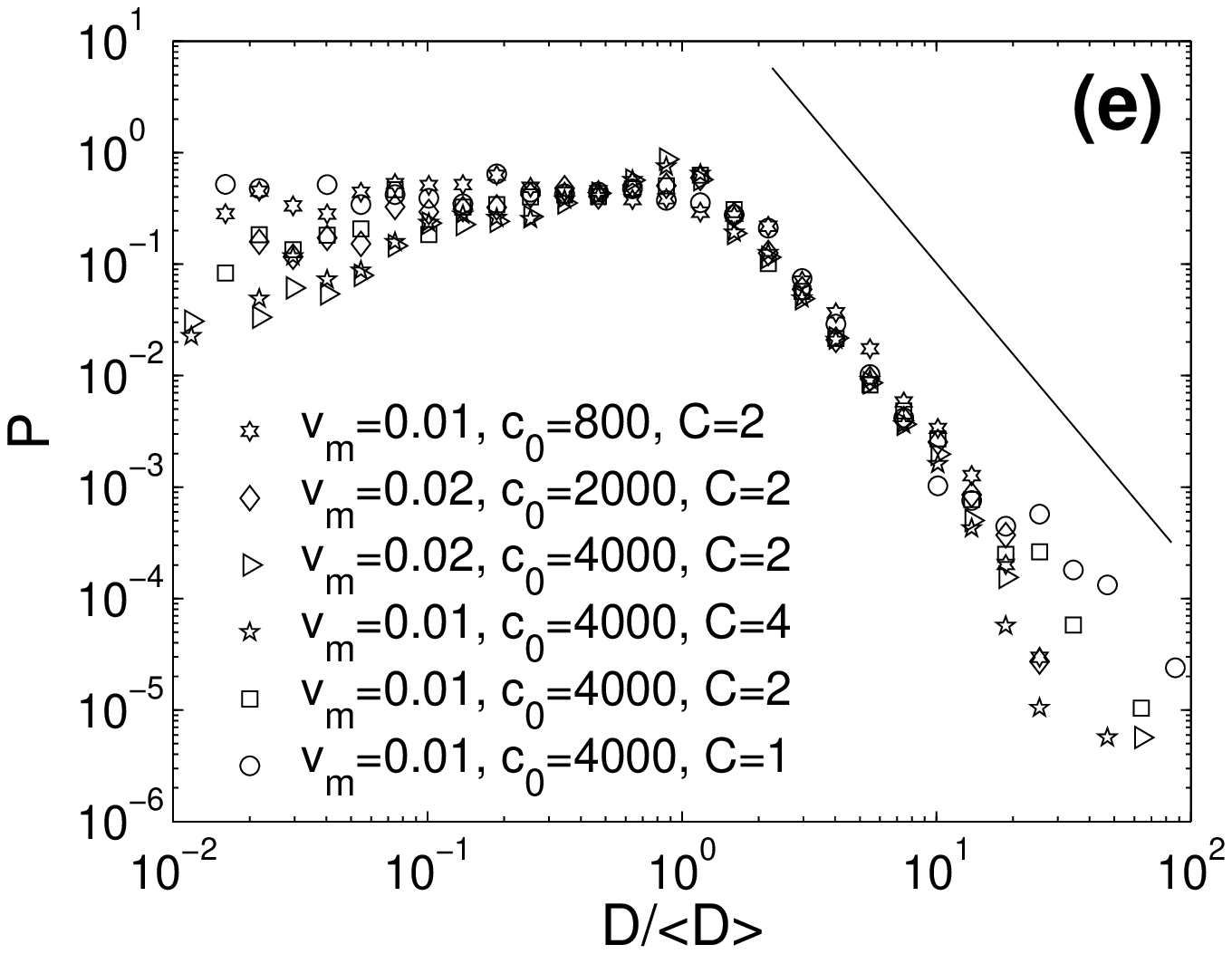}
\includegraphics[width=0.48\columnwidth]{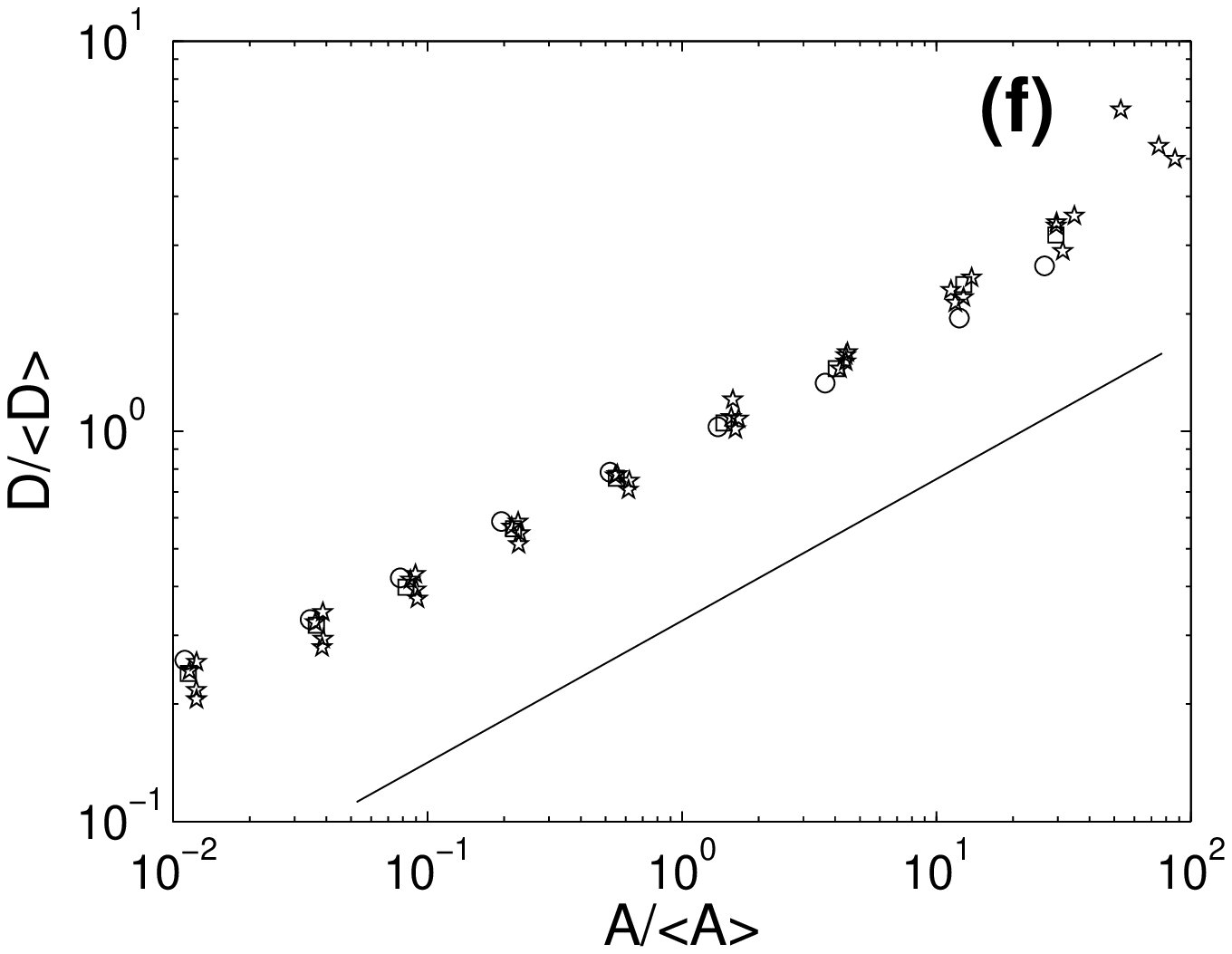}
\caption{(a) Typical gray scale map of the waiting time matrix $w(x,y)$ obtained from the solution of Eq. \ref{equ5} with $c_0=4000$ and $v_m=0.01$. (b) Spatial distribution of clusters corresponding to velocities $v(x,y) >C \bar{v}$ with $C=4$. The quantity $\bar{v}$ refers to the crack velocity averaged over both time and space within the steady regime. The clusters duration is given by the clusters color according to the colorscale given in inset. The distribution of the cluster area $A$, the scaling between the width $L_y$ and the length $L_x$ of clusters, the distribution of the cluster duration $D$, and the scaling between $D$ and $A$  are plotted in (c), (d), (e) and (f), respectively. The various symbols corresponds to various values of $v_m$, $c_0$ and $C$ as specified in the inset of Fig. e. The straight lines correspond to $P(A)\propto A^{-\tau_0}$, $L_y\propto L_x^{H}$, $P(D)\propto D^{-\alpha_>}$ and $D\propto A^\gamma$ with $\tau_0=-1.65$, $H=0.65\pm0.05$, $\alpha_>=-2.1$ and $\gamma=0.4$.}  
\end{center}
\label{fig2}
\end{figure}

{\it Spatially-averaged dynamics --} As for other critical systems, very different physical systems that belong to the same universality class will display similar scaling behaviours with same scaling exponents. In particular, it was shown that the Barkhausen noise \cite{Bertotti94_jap,Durin00_prl} accompanying the motion of domain wall driven by external magnetic field through soft disordered ferromagnets can be described by an equation with shape similar to Eq. \ref{equ5}. In this respect, we analyse the global crack front dynamics by computing the spatially averaged instantaneous velocity $v(t)=<\frac{\partial f}{\partial x}(x,t)>_x$  of the crack interface, as in \cite{Bertotti94_jap,Durin00_prl}. We observe once again a jerky dynamics really similar to Barkhausen noise \cite{Bertotti94_jap,Durin00_prl}. We then impose a given reference level $v_c=C\mathrm{max}(v)$ and define bursts as zones where $v(t)$ is above $v_c$. The duration $T$ of a given burst is defined as the interval within two successive intersections of $v(t)$ with $v_c$ while the size $S$ is defined as the integral of $v(t)$ between the same points. For the Barkhausen noise, this analysis leads to power-law distributions $P(T)\propto T^{-\alpha}$ and $P(S)\propto S^{-\tau}$ \cite{Bertotti94_jap,Durin00_prl}, and a power-law relation between the duration and size of the avalanches $T\propto S^a$, with critical exponents that can be predicted using Functional Renormalisation Group (FRG) calculations \cite{Durin00_prl,Ertas94_pre} leading to $\tau\approx 1.25$, $\alpha\approx 1.43$, and $a\approx 0.58$. We show on Fig. 3b, c and d. that all these scaling relations as well as the values of the exponents are found to be in good agreement with the ones observed here for the average interfacial crack growth.
\begin{figure}
\begin{center}
\includegraphics[height=0.38\columnwidth]{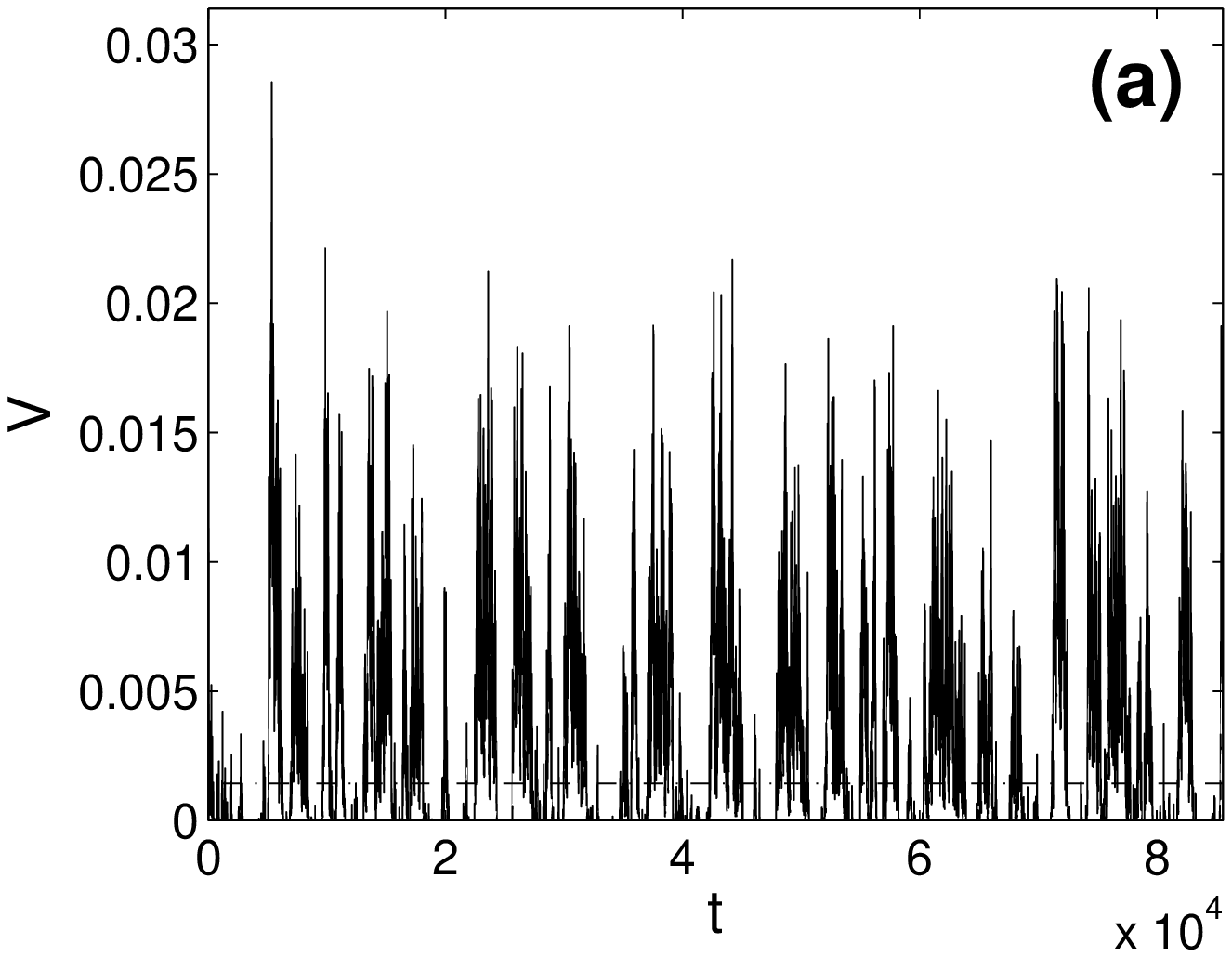}
\includegraphics[height=0.38\columnwidth]{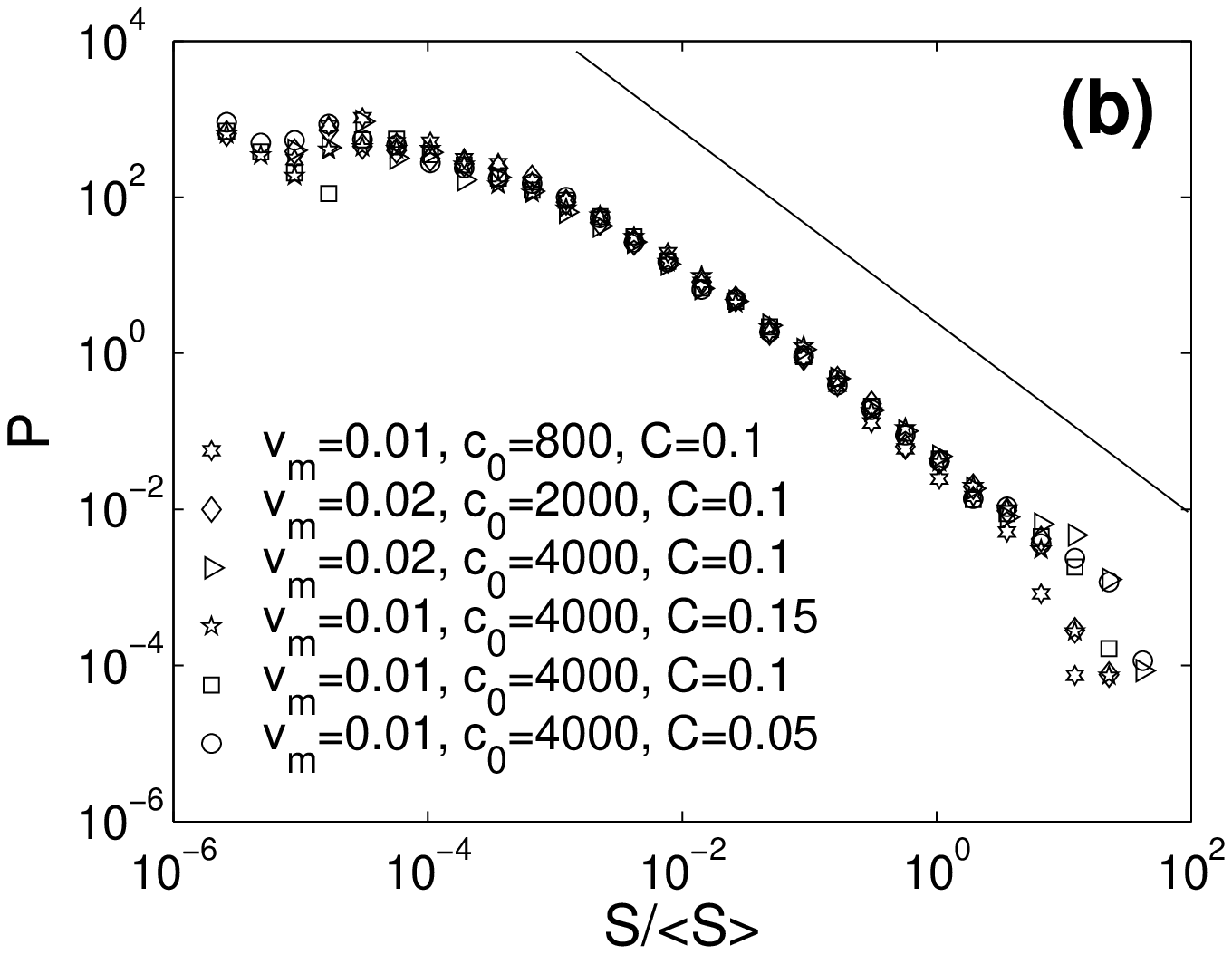}
\includegraphics[height=0.38\columnwidth]{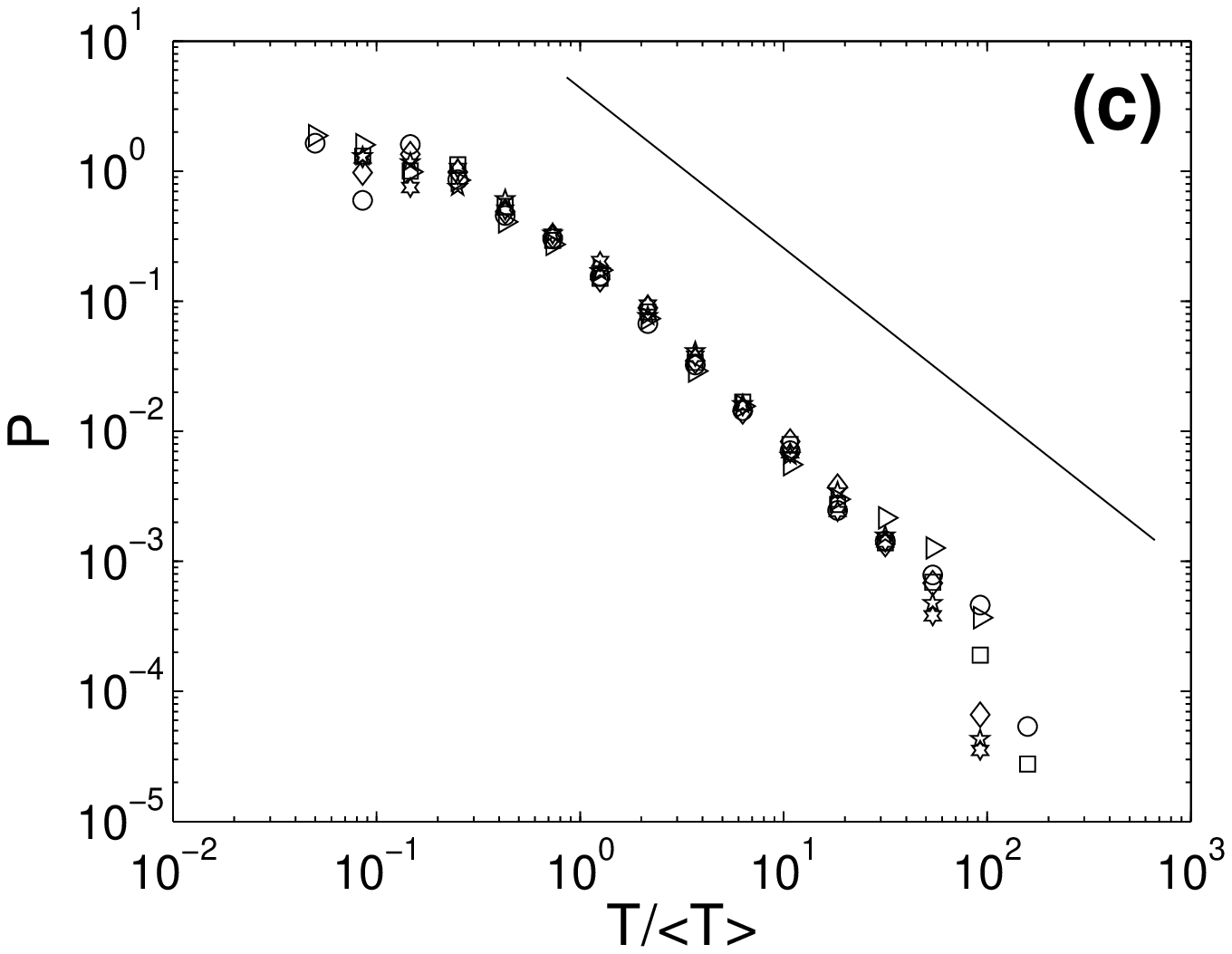}
\includegraphics[height=0.38\columnwidth]{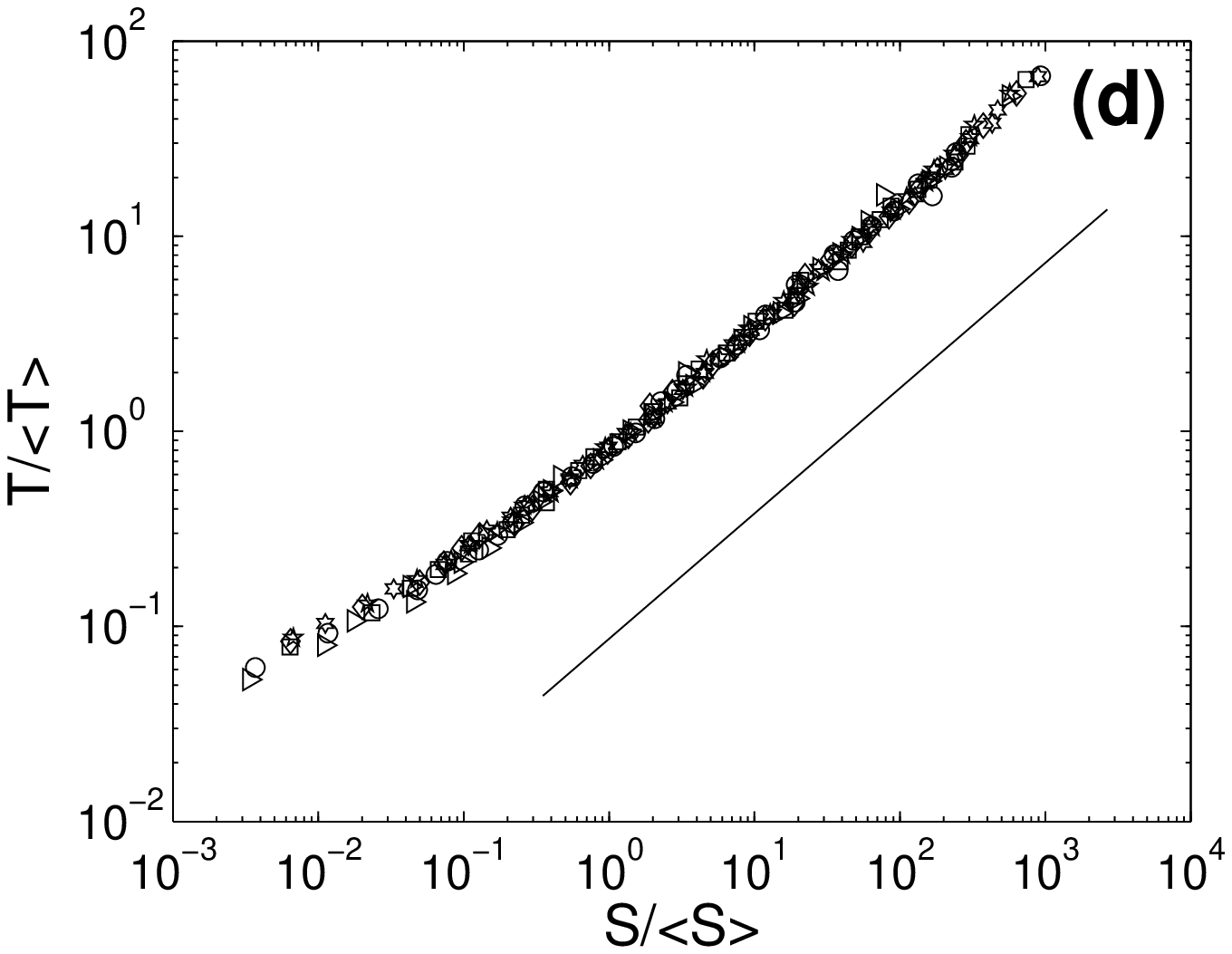}
\caption{(a): Typical time evolution of the spatially averaged crack front velocity $v(t)=<v(t,x)>_x$. The dash-dot horizontal line corresponds to $v_r=C \mathrm{max}(v)$ with $C=0.1$ used to define the bursts. Distribution of the normalized burst duration $T$, normalized burst size $S$, and the scaling of $T$ with $S$  are plotted in (b), (c) and (d), respectively. The various symbols corresponds to various values of $v_m$, $c_0$ and $C$ as specified in the inset of Fig. b. In these three graphs, the straight lines correspond to $P(T)\propto T^{-\alpha}$, $P(S)\propto S^{-\tau}$ and $T\propto S^a$ with the critical exponents $\alpha=1.43$, $\tau=1.25$, and $a=0.58$ predicted by FRG approach.}
\end{center}
\label{fig3}
\end{figure}

{\it Conclusion --} We have derived a description for planar crack growth in a disordered brittle material which succeeds to capture the statistics of the intermittent crackling dynamics recently observed experimentally \cite{Maloy06_prl}. In particular, we have shown that material failure appears as a critical system, where  the crack front progresses through scale-free avalanches signature of a dynamic depinning phase transition. As for other critical systems, microscopic and macroscopic details do not matter at large length and time scales and this simple Linear Elastic Stochastic description contains all the ingredients needed to capture the scaling statistical properties of more complex failure situations. This conjecture is strongly supported by a recent analysis \cite{Grob08_pageoph} that has shown that the laboratory experiment described here of crack gowth within Plexiglas under tension share many characteristics with geological faults -- although this latter results from external shear loading -- since they exhibit seismicity catalogs with similar satistical features: Time of occurrence, epicentre location and energy parameter. FRG calculations performed on a different system belonging to the same universality class, -- the motion of domain walls in disordered ferromagnets --, have been used to predict scaling laws for the statistics of the spatially averaged instantaneous crack velocity. Experimental checks of these predictions are currently under progress \cite{Santucci_cracklingexp}. Let us note finally that this description has been derived within the quasi-static approximation. To understand how to include the dynamic stress transfers through acoustic waves occurring as a {\it dynamically} growing crack is interacting with the material disorder \cite{dynamic} will represent interesting challenges for future investigations.

\begin{acknowledgments}
We thank K.J. M{\aa}l{\o}y, C. Guerra, C.L. Rountree, L. Barbier, E. Bouchaud, F. Célarié, P.-P. Cortet and L. Vanel for useful advices, helpful discussions, and  a critical reading of the manuscript. D.B. acknowledges funding from French ANR through grant No. ANR-05-JCJC-0088. S.S. was supported by the NFR Petromax Program No. 163472/S30.
\end{acknowledgments}


\begin{thebibliography}{35}
\expandafter\ifx\csname natexlab\endcsname\relax\def\natexlab#1{#1}\fi
\expandafter\ifx\csname bibnamefont\endcsname\relax
  \def\bibnamefont#1{#1}\fi
\expandafter\ifx\csname bibfnamefont\endcsname\relax
  \def\bibfnamefont#1{#1}\fi
\expandafter\ifx\csname citenamefont\endcsname\relax
  \def\citenamefont#1{#1}\fi
\expandafter\ifx\csname url\endcsname\relax
  \def\url#1{\texttt{#1}}\fi
\expandafter\ifx\csname urlprefix\endcsname\relax\def\urlprefix{URL }\fi
\providecommand{\bibinfo}[2]{#2}
\providecommand{\eprint}[2][]{\url{#2}}

\bibitem[{\citenamefont{Alava et~al.}(2006)\citenamefont{Alava, Nukala, and
  Zapperi}}]{Alava06_ap}
\bibinfo{author}{\bibfnamefont{M.~J.} \bibnamefont{Alava}},
  \bibinfo{author}{\bibfnamefont{P.~K.~V.~V.} \bibnamefont{Nukala}},
  \bibnamefont{and} \bibinfo{author}{\bibfnamefont{S.}~\bibnamefont{Zapperi}},
  \bibinfo{journal}{Adv. Phys.} \textbf{\bibinfo{volume}{55}},
  \bibinfo{pages}{349} (\bibinfo{year}{2006}).

\bibitem{Bouchaud97_jpcm}
see \bibinfo{author}{\bibfnamefont{E.}~\bibnamefont{Bouchaud}},
  \bibinfo{journal}{J. Phys. Condens. Matter} \textbf{\bibinfo{volume}{9}},
  \bibinfo{pages}{4319} (\bibinfo{year}{1997}) for a review and references therein.

\bibitem[{\citenamefont{Ponson et~al.}(2006{\natexlab{a}})\citenamefont{Ponson,
  Bonamy, and Bouchaud}}]{Ponson06_prl}
\bibinfo{author}{\bibfnamefont{L.}~\bibnamefont{Ponson}},
  \bibinfo{author}{\bibfnamefont{D.}~\bibnamefont{Bonamy}}, \bibnamefont{and}
  \bibinfo{author}{\bibfnamefont{E.}~\bibnamefont{Bouchaud}},
  \bibinfo{journal}{Phys. Rev. Lett.} \textbf{\bibinfo{volume}{96}},
  \bibinfo{pages}{035506} (\bibinfo{year}{2006}{\natexlab{a}}).
\bibinfo{author}{\bibfnamefont{L.}~\bibnamefont{Ponson}},
  \bibinfo{author}{\bibfnamefont{D.}~\bibnamefont{Bonamy}}, \bibnamefont{and}
  \bibinfo{author}{\bibfnamefont{L.}~\bibnamefont{Barbier}},
  \bibinfo{journal}{Phys. Rev. B} \textbf{\bibinfo{volume}{74}},
  \bibinfo{pages}{184205} (\bibinfo{year}{2006}{\natexlab{b}}).

\bibitem[{\citenamefont{Bonamy et~al.}(2006)\citenamefont{Bonamy, Ponson,
  Prades, Bouchaud, and Guillot}}]{Bonamy06_prl}
\bibinfo{author}{\bibfnamefont{D.}~\bibnamefont{Bonamy et al.}},
  \bibinfo{journal}{Phys. Rev. Lett.} \textbf{\bibinfo{volume}{97}},
  \bibinfo{pages}{135504} (\bibinfo{year}{2006}).

\bibitem{directcrackling}
\bibinfo{author}{\bibfnamefont{K.~J.} \bibnamefont{M{\aa}l{\o}y}} \bibnamefont{and}
  \bibinfo{author}{\bibfnamefont{J.}~\bibnamefont{Schmittbuhl}},
  \bibinfo{journal}{Phys. Rev. Lett.} \textbf{\bibinfo{volume}{87}},
  \bibinfo{pages}{105502} (\bibinfo{year}{2001});
\bibinfo{author}{\bibfnamefont{S.}~\bibnamefont{Santucci}},
  \bibinfo{author}{\bibfnamefont{L.}~\bibnamefont{Vanel}}, \bibnamefont{and}
  \bibinfo{author}{\bibfnamefont{S.}~\bibnamefont{Ciliberto}},
  \bibinfo{journal}{Phys. Rev. Lett.} \textbf{\bibinfo{volume}{93}},
  \bibinfo{pages}{095505} (\bibinfo{year}{2004});
\bibinfo{author}{\bibfnamefont{A.}~\bibnamefont{Marchenko}},
  \bibinfo{author}{\bibfnamefont{D.}~\bibnamefont{Fichou}},
  \bibinfo{author}{\bibfnamefont{D.}~\bibnamefont{Bonamy}}, \bibnamefont{and}
  \bibinfo{author}{\bibfnamefont{E.}~\bibnamefont{Bouchaud}},
  \bibinfo{journal}{Appl. Phys. Lett.} \textbf{\bibinfo{volume}{89}},
  \bibinfo{pages}{093124} (\bibinfo{year}{2006}).

\bibitem[{\citenamefont{Sethna et~al.}(2001)\citenamefont{Sethna, Dahmen, and
  Myers}}]{Sethna01_nature}
\bibinfo{author}{\bibfnamefont{J.~P.} \bibnamefont{Sethna}},
  \bibinfo{author}{\bibfnamefont{K.~A.} \bibnamefont{Dahmen}},
  \bibnamefont{and} \bibinfo{author}{\bibfnamefont{C.~R.} \bibnamefont{Myers}},
  \bibinfo{journal}{Nature} \textbf{\bibinfo{volume}{410}},
  \bibinfo{pages}{242} (\bibinfo{year}{2001}).

\bibitem{acoustic}
\bibinfo{author}{\bibfnamefont{A.}~\bibnamefont{Petri} et al.},
  \bibinfo{journal}{Phys. Rev. Lett.} \textbf{\bibinfo{volume}{73}},
  \bibinfo{pages}{3423} (\bibinfo{year}{1994});
\bibinfo{author}{\bibfnamefont{A.}~\bibnamefont{Garcimartin}},
  \bibinfo{author}{\bibfnamefont{A.}~\bibnamefont{Guarino}},
  \bibinfo{author}{\bibfnamefont{L.}~\bibnamefont{Bellon}}, \bibnamefont{and}
  \bibinfo{author}{\bibfnamefont{S.}~\bibnamefont{Ciliberto}},
  \bibinfo{journal}{Phys. Rev. Lett.} \textbf{\bibinfo{volume}{79}},
  \bibinfo{pages}{3202} (\bibinfo{year}{1997});
\bibinfo{author}{\bibfnamefont{J.}~\bibnamefont{Davidsen}},
  \bibinfo{author}{\bibfnamefont{S.}~\bibnamefont{Stanchits}},
  \bibnamefont{and} \bibinfo{author}{\bibfnamefont{G.}~\bibnamefont{Dresen}},
  \bibinfo{journal}{Phys. Rev. Lett.} \textbf{\bibinfo{volume}{98}},
  \bibinfo{pages}{125502} (\bibinfo{year}{2007});
\bibinfo{author}{\bibfnamefont{J.}~\bibnamefont{Koivisto}},
  \bibinfo{author}{\bibfnamefont{J.}~\bibnamefont{Rosti}}, \bibnamefont{and}
  \bibinfo{author}{\bibfnamefont{M.~J.} \bibnamefont{Alava}},
  \bibinfo{journal}{Phys. Rev. Lett.} \textbf{\bibinfo{volume}{99}},
  \bibinfo{pages}{145504} (\bibinfo{year}{2007}).

\bibitem{earthquakes}
\bibinfo{author}{\bibfnamefont{B.}~\bibnamefont{Gutenberg}} \bibnamefont{and}
  \bibinfo{author}{\bibfnamefont{C.~F.} \bibnamefont{Richter}},
  \bibinfo{title}{{\it Seismicity of the earth and associated phenomena.}}
  (\bibinfo{publisher}{Princeton University Press}, \bibinfo{year}{1954});
\bibinfo{author}{\bibfnamefont{A.}~\bibnamefont{Corral}},
  \bibinfo{journal}{Phys. Rev. Lett.} \textbf{\bibinfo{volume}{92}},
  \bibinfo{pages}{108501} (\bibinfo{year}{2004}).
  
\bibitem{FBM}
H.~E. Daniels, Proc. R. Acad. London A {\bf 183}, 405 (1945);
P.~C. Hemmer, A. Hansen and S. Pradhan, Lect. Notes Phys. {\bf 705}, 27 (2006);
F. Kun, F. Raisher, R.~C. Hidalgo and H.~J. Herrmann, Lect. Notes Phys. {\bf 705}, 57 (2006).

\bibitem{RFM}
L. de Arcangelis, S. Redner and H.~J. Herrmann, J. Physique Lett. {\bf 46}, 585 (1985);
S. Zapperi, P.K.V.V. Nukala and S. Simunovic, Phys. Rev. E {\bf 71}, 026106 (2005);
S. Zapperi, P.K.V.V. Nukala and S. Simunovic, Physica A {\bf 357}, 129 (2005).

\bibitem[{\citenamefont{Maloy et~al.}(2006)\citenamefont{Maloy, Santucci,
  Schmittbuhl, and Toussaint}}]{Maloy06_prl}
\bibinfo{author}{\bibfnamefont{K.~J.}~\bibnamefont{M{\aa}l{\o}y}},
  \bibinfo{author}{\bibfnamefont{S.}~\bibnamefont{Santucci}},
  \bibinfo{author}{\bibfnamefont{J.}~\bibnamefont{Schmittbuhl}},
  \bibnamefont{and}
  \bibinfo{author}{\bibfnamefont{R.}~\bibnamefont{Toussaint}},
  \bibinfo{journal}{Phys. Rev. Lett.} \textbf{\bibinfo{volume}{96}},
  \bibinfo{pages}{045501} (\bibinfo{year}{2006}).

\bibitem[{\citenamefont{Lawn}(1993)}]{Lawn93_book}
\bibinfo{author}{\bibfnamefont{B.}~\bibnamefont{Lawn}},
  \bibinfo{booktitle}{ {\it Fracture of brittle solids}}
  (\bibinfo{publisher}{Cambridge University Press}, 
 \bibinfo{year}{1993}).

\bibitem[{\citenamefont{Obreimoff}(1930)}]{Obreimoff30_rpsl}
\bibinfo{author}{\bibfnamefont{J.-W.} \bibnamefont{Obreimoff}},
  \bibinfo{journal}{Proc. Roy. Soc. London A}
  \textbf{\bibinfo{volume}{127}}, \bibinfo{pages}{290} (\bibinfo{year}{1930}).

\bibitem[{\citenamefont{Schmittbuhl et~al.}(1995)\citenamefont{Schmittbuhl,
  Roux, Vilotte, and Maloy}}]{Schmittbuhl95_prl}
\bibinfo{author}{\bibfnamefont{J.}~\bibnamefont{Schmittbuhl}},
  \bibinfo{author}{\bibfnamefont{S.}~\bibnamefont{Roux}},
  \bibinfo{author}{\bibfnamefont{J.~P.} \bibnamefont{Vilotte}},
  \bibnamefont{and} \bibinfo{author}{\bibfnamefont{K.~J.} \bibnamefont{M{\aa}l{\o}y}},
  \bibinfo{journal}{Phys. Rev. Lett.} \textbf{\bibinfo{volume}{74}},
  \bibinfo{pages}{1787} (\bibinfo{year}{1995});
  \bibinfo{author}{\bibfnamefont{S.}~\bibnamefont{Ramanathan}},
  \bibinfo{author}{\bibfnamefont{D.}~\bibnamefont{Ertas}}, \bibnamefont{and}
  \bibinfo{author}{\bibfnamefont{D.~S.} \bibnamefont{Fisher}},
  \bibinfo{journal}{Phys. Rev. Lett.} \textbf{\bibinfo{volume}{79}},
  \bibinfo{pages}{873} (\bibinfo{year}{1997}).

\bibitem[{\citenamefont{Gao and Rice}(1989)}]{Gao89_jam}
\bibinfo{author}{\bibfnamefont{H.}~\bibnamefont{Gao}} \bibnamefont{and}
  \bibinfo{author}{\bibfnamefont{J.~R.} \bibnamefont{Rice}},
  \bibinfo{journal}{J. Appl. Mech.} \textbf{\bibinfo{volume}{56}},
  \bibinfo{pages}{828} (\bibinfo{year}{1989}).
  
\bibitem{roughnessexponent}
A. Tanguy, M. Gounelle and S. Roux, Phys. Rev. E {\bf 58}, 1577 (1998);
A. Rosso and W. Krauth, Phys. Rev. E {\bf 65}, 025101(R) (2002);

\bibitem{Santucciroughness08}
S. Santucci et al to be submitted (2008). 


\bibitem[{\citenamefont{Bertotti et~al.}(1994)\citenamefont{Bertotti, G, and
  Magni}}]{Bertotti94_jap}
\bibinfo{author}{\bibfnamefont{G.}~\bibnamefont{Bertotti}},
  \bibinfo{author}{\bibfnamefont{G.~D.} \bibnamefont{G}}, \bibnamefont{and}
  \bibinfo{author}{\bibfnamefont{A.}~\bibnamefont{Magni}},
  \bibinfo{journal}{J. Appl. Phys.} \textbf{\bibinfo{volume}{75}},
  \bibinfo{pages}{5490} (\bibinfo{year}{1994});
\bibinfo{author}{\bibfnamefont{S.}~\bibnamefont{Zapperi}},
  \bibinfo{author}{\bibfnamefont{P.}~\bibnamefont{Cizeau}},
  \bibinfo{author}{\bibfnamefont{G.}~\bibnamefont{Durin}}, \bibnamefont{and}
  \bibinfo{author}{\bibfnamefont{H.~E.} \bibnamefont{Stanley}},
  \bibinfo{journal}{Phys. Rev. B} \textbf{\bibinfo{volume}{58}},
  \bibinfo{pages}{6353} (\bibinfo{year}{1998}).

\bibitem[{\citenamefont{Durin and Zapperi}(2000)}]{Durin00_prl}
\bibinfo{author}{\bibfnamefont{G.}~\bibnamefont{Durin}} \bibnamefont{,}
  \bibinfo{author}{\bibfnamefont{S.}~\bibnamefont{Zapperi}},
  \bibinfo{journal}{Phys. Rev. Lett.} \textbf{\bibinfo{volume}{84}},
  \bibinfo{pages}{4705} (\bibinfo{year}{2000}).

\bibitem[{\citenamefont{Ertas and Kardar}(1994)}]{Ertas94_pre}
\bibinfo{author}{\bibfnamefont{D.}~\bibnamefont{Ertas}} \bibnamefont{and}
  \bibinfo{author}{\bibfnamefont{M.}~\bibnamefont{Kardar}},
  \bibinfo{journal}{Phys. Rev. E} \textbf{\bibinfo{volume}{49}},
  \bibinfo{pages}{R2532} (\bibinfo{year}{1994}).

\bibitem[{\citenamefont{Rolley et~al.}(1998)\citenamefont{Rolley, Guthmann,
  Gombrowicz, and Repain}}]{Rolley98_prl}
\bibinfo{author}{\bibfnamefont{E.}~\bibnamefont{Rolley}},
  \bibinfo{author}{\bibfnamefont{C.}~\bibnamefont{Guthmann}},
  \bibinfo{author}{\bibfnamefont{R.}~\bibnamefont{Gombrowicz}},
  \bibnamefont{and} \bibinfo{author}{\bibfnamefont{V.}~\bibnamefont{Repain}},
  \bibinfo{journal}{Phys. Rev. Lett.} \textbf{\bibinfo{volume}{80}},
  \bibinfo{pages}{2865} (\bibinfo{year}{1998}).

\bibitem{NarayanNattermannKardar}
O. Narayan and D. S. Fisher, Phys. Rev. B {\bf 48}, 7030 (1993);
H. Leschhorn, T. Nattermann, S. Stepanow and L.~H. Tang, Ann. Physik {\bf 509}, 1–34 (1997);

\bibitem[{\citenamefont{Bak et~al.}(1987)\citenamefont{Bak, Tang, and
  Wiesenfield}}]{Bak87_prl}
\bibinfo{author}{\bibfnamefont{P.}~\bibnamefont{Bak}},
  \bibinfo{author}{\bibfnamefont{C.}~\bibnamefont{Tang}}, \bibnamefont{and}
  \bibinfo{author}{\bibfnamefont{K.}~\bibnamefont{Wiesenfeld}},
  \bibinfo{journal}{Phys. Rev. Lett.} \textbf{\bibinfo{volume}{59}},
  \bibinfo{pages}{381} (\bibinfo{year}{1987}).

\bibitem{Santucci_cracklingexp}
S. Santucci et al, in preparation (2008). 

\bibitem{Grob08_pageoph}
M. Grob et al., to appear in PAGEOPH (2008).

\bibitem{dynamic}
K. Ravi-Chandar and W. Knauss, Int. J. Frac. {\bf 26}, 189 (1984);
E. Sharon, E. Cohen and J. Fineberg, Nature {\bf 410}, 68 (2001);
D. Bonamy and K. Ravi-Chandar, Phys. Rev. Lett. {\bf 91}, 235502 (2003);
D. Bonamy and K. Ravi-Chandar, Int. J. Frac. {\bf 134}, 1 (2005).

\end{thebibliography}
\end{document}